\documentclass[conference]{IEEEtran}
\IEEEoverridecommandlockouts
\usepackage{cite}
\usepackage{amsmath,amssymb,amsfonts}
\usepackage{algorithmic}
\usepackage{graphicx}
\usepackage{textcomp}
\usepackage{xcolor}
\usepackage{gensymb}
\usepackage{amsmath}
\usepackage{algorithm}
\usepackage{hyperref}
\usepackage{capt-of}
\usepackage{cuted}
\usepackage{multirow}
\def\BibTeX{{\rm B\kern-.05em{\sc i\kern-.025em b}\kern-.08em
T\kern-.1667em\lower.7ex\hbox{E}\kern-.125em}}
\begin{document}

\title{Multi-Feeder Restoration using Multi-Microgrid Formation and Management
    % \thanks{This work was supported by American Public Power Association's DEED program.}
}

\author{
    Valliappan Muthukaruppan, Rongxing Hu, Ashwin Shirsat,\\ Mesut Baran,
    Ning Lu, Wenyuan Tang, and David Lubkeman \thanks{ The authors are with
      Department of Electrical and Computer Engineering, North Carolina State
      University, Raleigh, NC 27695 USA. This material is based upon work
      supported by U.S. Depart- ment of Energy’s Office of Energy Efficiency and
  Renewable Energy (EERE) under Solar Energy Technologies Office Award Number
DE-EE0008770.} }

\iffalse
    \author{\IEEEauthorblockN{Valliappan Muthukaruppan, \\Ning Lu, Mesut Baran}
        \IEEEauthorblockA{\textit{North Carolina State University} \\
        \textit{vmuthuk2, nlu2, baran @ncsu.edu}}
        \and
        \IEEEauthorblockN{Edmond Miller P.E., Matthew Makdad P.E.}
        \IEEEauthorblockA{\textit{New River Light and Power (NRLP)}\\
        \textit{millerec1, makdadmj @appstate.edu}}
        \and
        \IEEEauthorblockN{PJ Rehm}
        \IEEEauthorblockA{\textit{ElectriCities}\\
        \textit{prehm@electricities.org}}
    }
\fi

\maketitle

\begin{abstract}
  This papers highlights the benefit of coordinating resources on mulitple
  active distribution feeders during severe long duration outages through
  multi-microgrid formation. A graph-theory based multi-microgrid formation
  algorithm is developed which is agnostic of the underlying energy management
  scheme of the microgrids and solved in a rolling horizon fashion. The
  algorithm is then enhanced to handle multiple feeders where formation of long
  laterals needs to be avoided due to potential voltage control issues in
  distribution systems. The algorithm is evaluated on a synthetic two feeder
  system derived from interconnecting two IEEE 123 node system. The results
  indicate increased service to loads in the system and better utilization of
  renewable resources.
\end{abstract}


\section{Introduction}\label{sec:Introduction}
Forming a microgrid within a distribution system following a significant
weather-related outage is gaining prominence as a practical approach to enhance
resilience at the distribution level. This approach becomes particularly
appealing when the distribution system incorporates substantial amount of
distributed photovoltaic (PV) generation \cite{Labonte_2023}. Nevertheless, the
management of these microgrids at the feeder level or involving multiple
feeders presents several challenges. These challenges encompass constrained
resources for rapid deployment on the feeders and limited capabilities for
real-time monitoring and control within the distribution system \cite{SaeedEtAl_2021}.

In the presence of significant amount of renewable resources like rooftop solar
and large service territory, it is challenging for a microgrid with low inertial
resources to maintain power balance and support service over a long period of
time \cite{Pratt_2021}. Hence, it is beneficial to manage multiple microgrids
together. When it comes to multi-microgrid management during power system
restoration two methods have found prominence in literature, networked
microgrids \cite{ArifEtAl_2016} and dynamic multi-microgrid formation
\cite{IgderEtAl_2023}.

There are significant challenges with managing networked microgrids due to
complicated control requirements \cite{IslamEtAl_2021}. Hence, dynamic
multi-microgrid is a good solution especially for Utilities managing microgrids
owned by different stakeholders during extreme weather events. The control
architecture is simple and the boundary of the microgrids can be easily
controlled by Utilities. In existing literature, this problem is solved as
a critical service restoration problem where critical loads are routed to
existing microgrid resources through numerous circuit switches
\cite{XuEtAl_2018}. The issue with this type of restoration approach is that the
topology of microgrid remains same throughout the restoration process once the
critical loads are energized, which may not be optimal for all operating
conditions.

Another approach with multi-microgrid formation is to use mobile devices and
dispatch them from one location to another during the restoration as the system
condition changes \cite{LeiEtAl_2018}. The focus here is the routing of mobile
devices to designated load zones which is solved as a planning problem. Even
though this seems like a viable solution, the time taken to interconnect mobile
devices with distribution system is ignored which can hinder timely resotration
of loads and increase their downtime as well.

Current works in literature do not consider the challenging and realistic
conditions of a distribution system such as limited microgrid resources, limited
controllability, and significantly high solar penetration in the problem
formulation. Furthermore, none of the existing work have considered
a multi-feeder problem and the inherent issues in the setup.

This paper aims at developing a comprehensive restoration strategy for multiple
feeders using multi-microgrid formation. The main contribution of the proposed
method are listed below:
\begin{itemize}
  \item Multi-feeder restoration through formation of multiple-microgrids has
    a risk of leading to long laterals which can lead to voltage issues. We
    consider a special formulation of multi-microgrid problem that addresses
    this issue by providing DSO with a control on length of laterals that can be
    formed during the microgrid formation. Furthermore, the proposed formulation
    is agnostic of underlying energy management scheme of the microgrids and
    significantly improves the restoration process by coordinating the resources
    in different feeders.
  \item Realistic distribution sytem operating conditions are considered:
    limited load and PV visibility and controllability, and limited controllable
    swtiches on the feeder. The load and PV are assumed to be controlled only in
    zones formed by the existing circuit switches in the network.
\end{itemize}

The rest of the paper is organized as follows: sec-\ref{sec:problemformulation}
introduces the multi-feeder multi-microgrid formation problem and the interface
with the microgrid energy management schemes, sec-\ref{sec:Results} illustrates
the performance of proposed scheme with a case study using synthetic
multi-feeder sytem derived from two IEEE 123 node feeders and multi-day
operating conditions based on field data.

\section{Problem Formulation}%
\label{sec:problemformulation}
\subsection{Microgrid Formation}\label{sec:microgridformation}
The underlying multi-microgrid formation problem is a form of graph splitting
problem which can be solved using the single commodity flow method from graph
theory \cite{DingEtAl_2017}. Let $G=(V, E)$ be a connected and undirected graph
with set of vertices and edges denoted by $V$ and $E$ respectively representing
the distribution network under outage. Let the binary decision variables
$y_{ij}$ represent the status of circuit switches connecting load groups
\(i\),\(j\). If the switch is open,  $y_{ij} = 0$; otherwise, $y_{ij} = 1$. We
assume that there will be only one grid forming source per microgrid, thus the
number of microgrids is equal to the number of master control units. Which
means, $\mid \Pi \mid$  number of microgrids will be formed where $\Pi$ is the
set of grid forming resources in the network.

Objective function is given in (\ref{eq:mmgobj}) where the first term minimizes
the total load shedding in the network during microgrid formation where \(
D_j^{max} \) indicates the total connected load in load zone \( j \) and \(
\mathcal{N}^d \) indicates the total number of load zones in the network. The
second term minimizes the fictitious flow \( F_{ij} \) in the lines with higher
priority to lines connected directly to critical loads. The larger weight on
incoming flows to critical loads ensures that they are closer to the grid
forming sources of the microgrid thereby ensuring higher probability of service
during the energy management phase.

\begin{equation}\label{eq:mmgobj}
  \min_y \sum_{j=1}^{ \mid \mathcal{N}^d \mid } (D_j^{max} - D_j) + \sum_{ij\in E} w_j\mid F_{ij} \mid
\end{equation}

Let, $\mathcal{N}_p$ be the index set of number of microgrids to be formed. The
binary variable $x_{i,k} = 1$ indicates that load group \( i \) belongs to
microrgrid \( k \). Equation (\ref{eq:mmg1}) ensures that each load group
belongs to only one microgrid. While equations (\ref{eq:mmg2}-\ref{eq:mmg3})
ensures that nodes \( (f, k) \) connected to a closed switch \( l \) belong to
the same microgrid \( k \).

\begin{align}
  \sum_{i\in V} x_{i,k} &= 1 &\quad \forall k \in 1, \ldots, \mathcal{N}_p\label{eq:mmg1}\\
  y_l &= \sum_{k \in 1, \ldots, \mathcal{N}_p} z_{l, k} \quad &\forall l \in E\label{eq:mmg2}\\
  z_{l,k} &= x_{f,k}x_{t,k} \quad &\forall l\in E, (f,t) \in E(l)\label{eq:mmg3}
  % &z_{l,k} \le x_{f,k}; \quad z_{l,k} \le x_{t,k}; &\quad \forall l\in E, (f,t)
  % \in E(l)\label{eq:mmg3}\\
  % &z_{l,k} \le x_{f,k} + x_{l,k} - 1 &\quad \forall l\in E, (f,t) \in E(l)\label{eq:mmg3a}
\end{align}

For all \( l \in E \) and \( (f, t) \in E(l) \), equation (\ref{eq:mmg3a}) is
the McCormick Linearization of (\ref{eq:mmg3}).
\iffalse
\begin{equation}\label{eq:mmg3a}
  \begin{align*}
  &z_{l,k} \le x_{f,k}; \\
  &z_{l,k} \le x_{t,k}; \\
  &z_{l,k} \le x_{f,k} + x_{l,k} - 1
  \end{align*}
\end{equation}
\fi
\begin{align}
  z_{l,k} &\le x_{f,k};\notag\\
  z_{l,k} &\le x_{t,k};\label{eq:mmg3a}\\
  z_{l,k} &\le x_{f,k} + x_{l,k} - 1\notag
\end{align}

For all \( (f(l), t(l)) \in E(l) \), \( l \in E \), and \( i \in V \),
(\ref{eq:mmg4})-(\ref{eq:mmg8}) define the DC power flow equations with limits
on the line flow, power generation, and demand. Where, \( T_l \) is the net load
flowing through line \( l \) connecting load zones \( f(l) \text{ and } t(l) \).
\( P \) and \( D \) indicate the total PV generation and load in the individual
load zones. \( y_l \) indicates the status of switch \( l \).

\begin{align}
  \sum_{l:f(l)\rightarrow t(l)} T_l - \sum_{l:t(l)\rightarrow f(l)} T_l &= P_{f(l)} - D_{f(l)} \label{eq:mmg4}\\
  -T_l^{min} y_l &\le T_l \le T_l^{max} y_l \label{eq:mmg5}\\
  -M(1-y_l) &\le T_l \le M(1-y_l) \label{eq:mmg6}\\
  P_i^{min} &\le P_i \le P_i^{max} \label{eq:mmg7}\\
  D_i^{min} &\le D_i \le D_i^{max} \label{eq:mmg8}
\end{align}

Equation (\ref{eq:mmg9}) ensures radiality in the network by determining the
total number of closed switches possible. Here, $ \mid V \mid $ is the number of
nodes, $ \mid \Pi \mid $ is the number of microgrid resources available in the
system, and $ \mid \mathcal{R} \mid $ is the number of load islands. Load
islands are load groups that cannot be connected to any microgrid resources due
to existing faults in the system.

\begin{equation}\label{eq:mmg9}
  \sum_{ij\in E} y_{ij} =  \mid V \mid -  \mid \Pi \mid -  \mid \mathcal{R}  \mid
\end{equation}

To ensure the individual microgrid connectivity via mathematical programming
formulations, the single commodity flow method is employed as shown in equations
(\ref{eq:mmg10}) - (\ref{eq:mmg15}). The equations are similar to the DC power
flow equations but all nodes are assumed to inject a load of value 1 and the
sources are indicated by \( W \)

\begin{align}
  \sum_{s\in\delta(j)} F_{js} - \sum_{i\in\pi(j)} F_{ij} &= 1 &\quad \forall j \in V/\Pi\label{eq:mmg10}\\
  \sum_{s\in\delta(j)} F_{js} - \sum_{i\in\pi(j)} F_{ij} &= W_j &\quad \forall j \in \Pi\label{eq:mmg11}\\
  -My_{ij} \le F_{ij} \le & My_{ij} &\quad \forall ij \in E\label{eq:mmg12}\\
  -M(2-y_{ij}) \le F_{ij} \le & M(2-y_{ij}) &\quad \forall ij \in E\label{eq:mmg13}\\
  W_j \le & 1 &\quad \forall j \in \Pi\label{eq:mmg14}\\
  F_{ij} \ge & n^{min} &\quad \forall i \in \mathcal{N}_{GFM} \label{eq:mmg15}
\end{align}

\begin{figure}[htbp]
  \centering
  \includegraphics[width=3.5in]{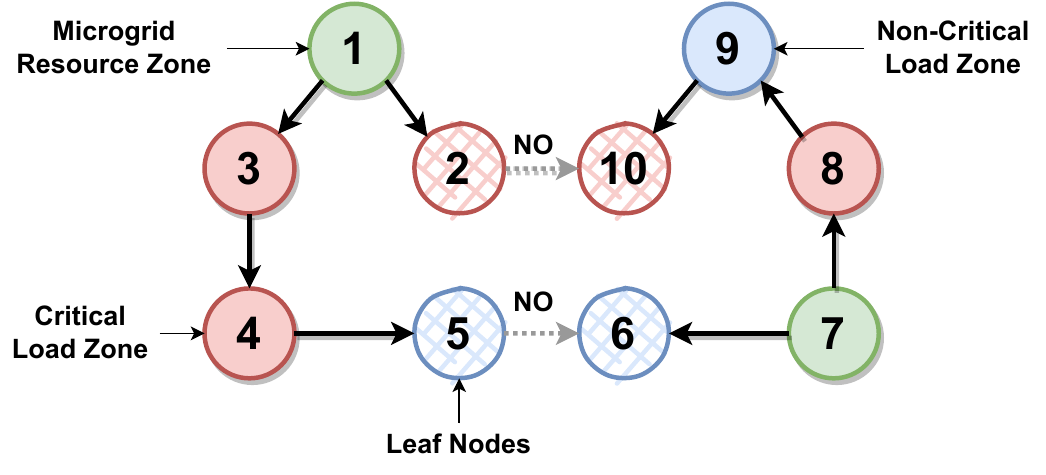}
  \caption{Graphical representation of a two feeder restoration problem}
  \label{fig:mmgtopo}
\end{figure}

Consider the simplified graphical representation of a simple multi-feeder
restoration example as shown in fig. \ref{fig:mmgtopo}. Nodes 1 through 5 belong
to feeder-1 and 6 through 10 belong to feeder-2. There are two normally open
switches interconnecting the nearby feeders. When restoring multiple feeders
using multiple microgrids, the length of the radial network inside each
microgrid needs to be controlled since the underlying distribution system
voltage regulation may not be capable of handling the different microgrid
topologies. Hence, it is critical to limit the boundary of the microgrids.

To avoid creating long radial networks, we modify the single commodity flow
method such that only leaf nodes (feeder-1: nodes 2 and 5; feeder-2: nodes 6 and
10) will be exchanged between the microgrids during restoration. Leaf nodes are
the nodes connected to the normally open switches in the network. This strategy
is implemented using constraint (\ref{eq:mmg15}) where we control the fictious
flow on the lines directly connected to the grid forming sources (root of the
graph) in each microgrid.

\begin{figure}[htbp]
  \centering
  \includegraphics[width=2.5in]{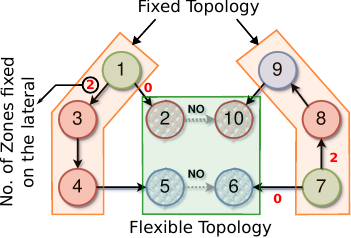}
  \caption{Fixed and Flexible Topology Classification to avoid voltage issues
  caused by long radial networks because of picking up too many load groups from
  adjacent feeder.}
  \label{fig:mmgtopoex}
\end{figure}

For example, in figure \ref{fig:mmgtopoex} the left lateral of node-1 with grid
forming resources (GFM) in MG-1 is restricted to two such that load groups 3 and
4 are always connected to MG-1 while node-5 can be interchanged with the
neighboring microgrids. Since node-2 is directly connected to the GFM in MG-1
the fictious flow on the right branch of GFM is set to zero. This approach can
be extended any number of feeders with interconnecting switches and different
topologies.

\subsection{Energy Management}
The advantage of the proposed microgrid formation problem is that it is agnostic
of the underlying energy management scheme (EMS), each microgrid can use its own
EMS implemented through its microgrid controller. No modification to the
underlying control architecture is necessary. In this paper we use our
previously proposed rolling horizon based two-stage hierarchical EMS for each
estabilished microgrid \cite{MuthukaruppanEtAl_2022}.

The decision variables from multi-microgrid formation problem is the optimal
topology of individual microgrids which will be relayed to the microgrid
controllers. The microgrid controllers will then manage the resources and load
groups in their assigned topology until next topology change by multi-microgrid
formation module.

\begin{figure}[htbp]
  \centering
  \includegraphics[width=3.5in]{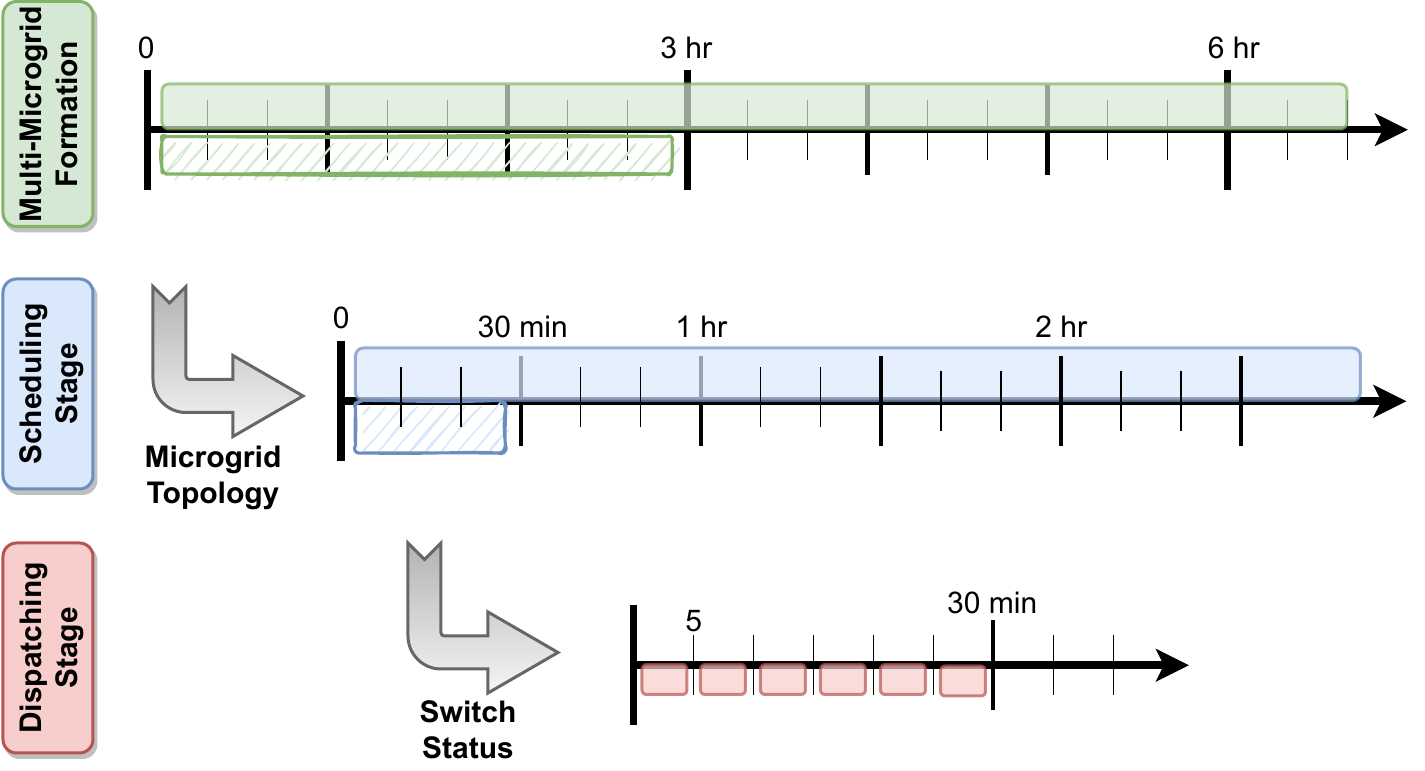}
  \caption{Timeline of Different problems and their coordination}
  \label{fig:mmgtimeline}
\end{figure}

The horizon, time step, and coordination of the various modules in the
multi-microgrid formation and management problem is shown in figure
\ref{fig:mmgtimeline}. The multi-microgrid formation module is also formulated
as a rolling horizon problem to avoid frequent topology changes in the
microgrids. Note that equations in section-\ref{sec:microgridformation} are
defined for a single time step for ease of explanation but the problem is solved
for a whole horizon with multiple timesteps.

Eventhough the multi-microgrid fomration module determines the optimal topology
of each microgrid, the decision to supply the load groups within each microgrid
is still determined by the individual microgrid EMS system depending on the
availability of the resources.

\section{Results}\label{sec:Results}
\begin{figure}[htbp]
  \centering
  \includegraphics[width=3.5in]{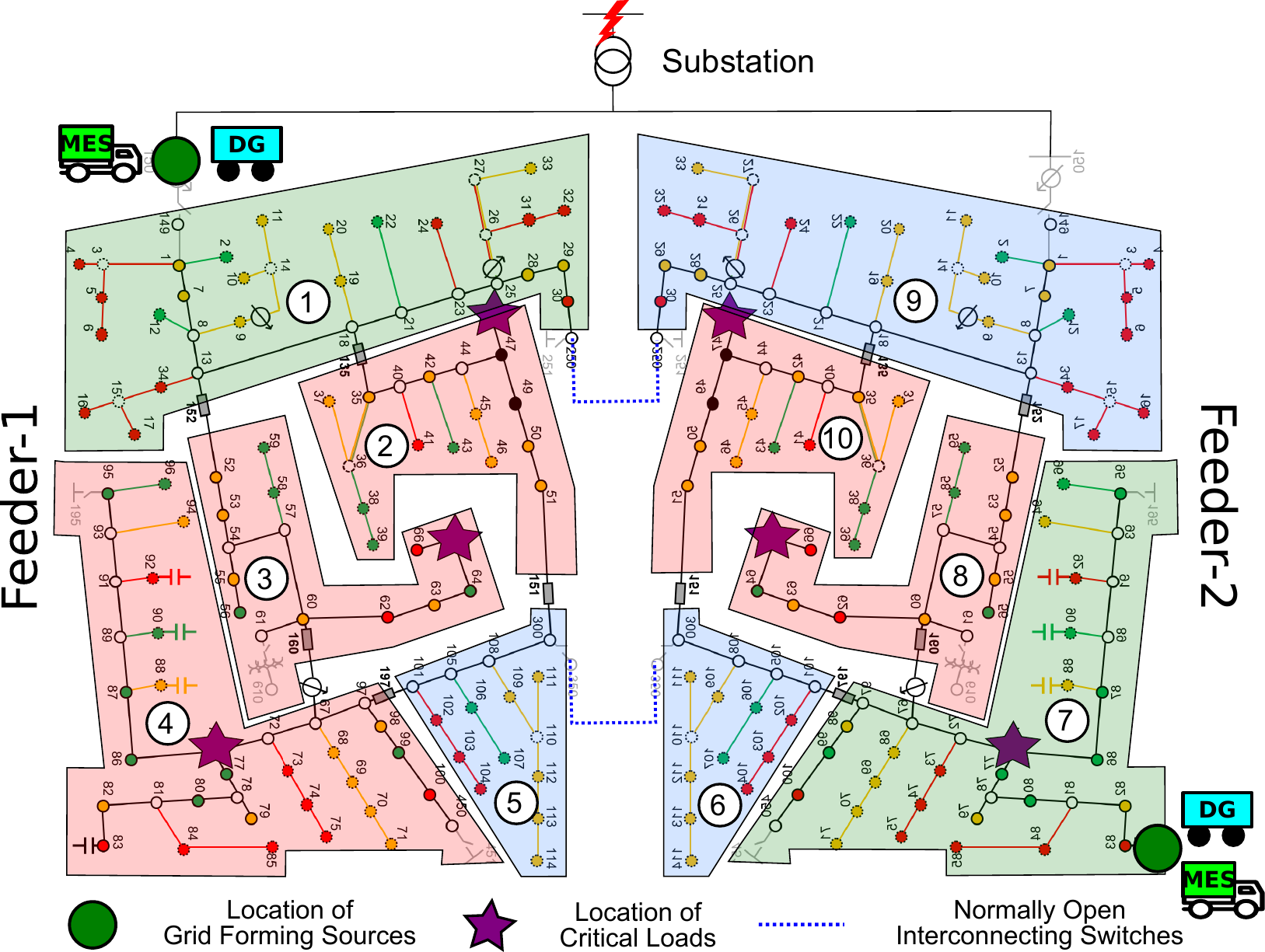}
  \caption{Multi-feeder test system developed from two IEEE 123 node feeders}
  \label{fig:mmgcasestudy}
\end{figure}

The test system considered for evaluating the proposed algorithm is shown in
figure~\ref{fig:mmgcasestudy}. We have used two IEEE 123 node system which are
assumed to be supplied from same substation to develop the multi-feeder test
case. The location of microgrid resources on feeder-1 is at the substation while
in feeder-2 they are towards the end of the feeder. These locations are
pre-determined and fixed throughout the restoration. Two interconnecting
switches which are normally open at nodes 250 and 350 interconnect the two IEEE
123 feeders. All load groups defined by the shaded polygons have significant
penetration of distributed behind-the-meter PV. There are 3 critical loads per
feeder highlighted as purple stars in the figure. The graphical representation
of this system is shown in figure~\ref{fig:mmgtopo}. The rating of the resources
are highlighted in table~\ref{tab:mmgresources}. The horizon and time step of
the EMS and MMG formation module are highlighted in table~\ref{tab:mmghorizon}.

\begin{table}[htpb]
  \centering
  \caption{Rating of resources in the multi-feeder test case}
  \label{tab:mmgresources}
  \begin{tabular}{l|c|c}
  \hline
  \textbf{Resources} & \textbf{Feeder-1} & \textbf{Feeder-2}\\
  \hline
  \hline
  Mobile Energy Storage & 3 MW/ 12 MWh & 2 MW/ 8 MWh\\
  Diesel Generator      & 4 MW         & 4 MW       \\
  Total PV              & 4 MW         & 4 MW       \\
  Peak Load (Day-1)     & 3.5 MW       & 3 MW       \\
  Peak Load (Day-2)     & 3 MW         & 2 MW       \\
  \hline
  \end{tabular}
\end{table}

\begin{table}[htpb]
  \centering
  \caption{Horizon and Time step of different problems in MMG restoration}
  \label{tab:mmghorizon}
  \begin{tabular}{l|c|c}
  \hline
  \textbf{Problem} & \textbf{Horzion} & \textbf{Time Step}\\
  \hline
  \hline
  Multi-Microgrid Formation & 24 hour & 3 hour\\
  Scheduling Stage-1 & 24 hour & 30 minutes\\
  Dispatching Stage-2 & 30 minutes & 5 minute\\
  \hline
  \end{tabular}
\end{table}

\begin{figure}[htbp]
  \centering
  \includegraphics[width=3.5in]{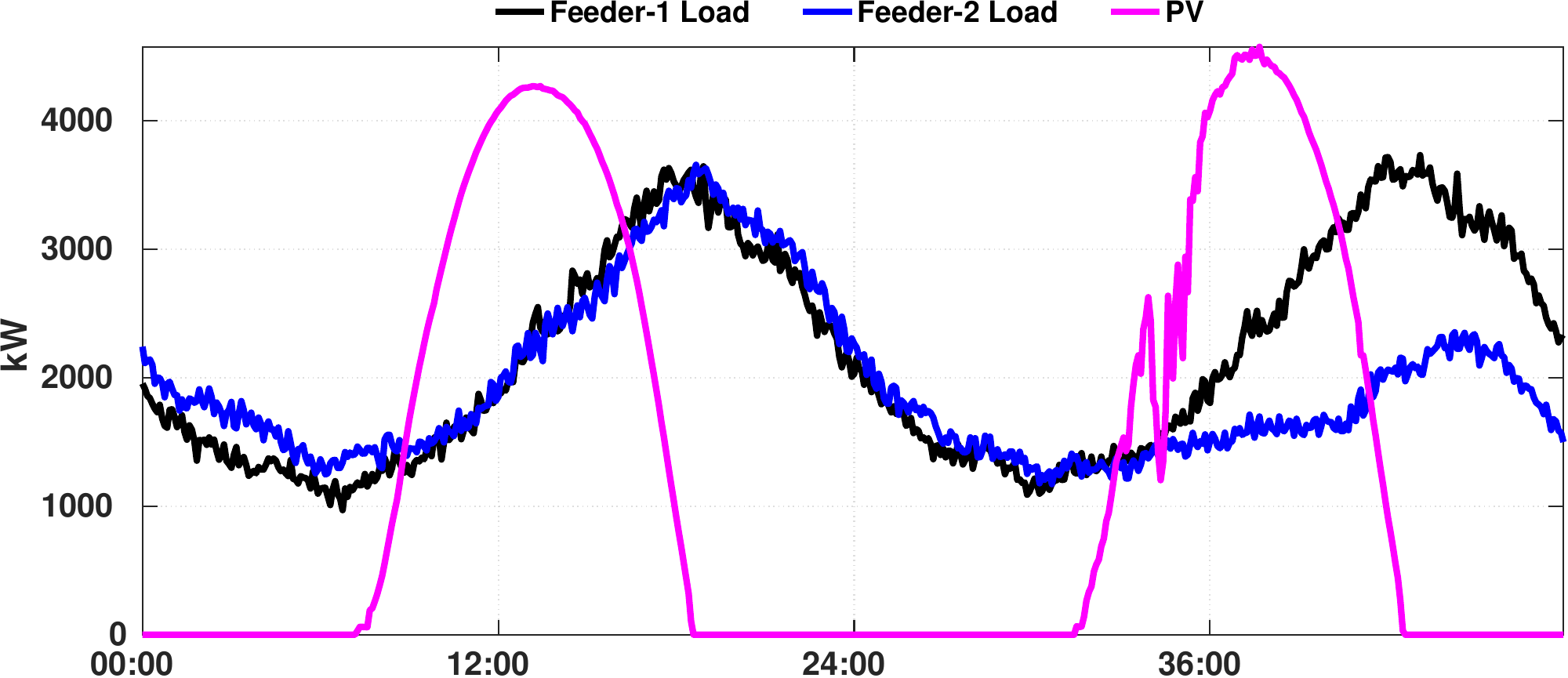}
  \caption{Total Load and PV on feeder-1 and feeder-2}
  \label{fig:mmgldpv}
\end{figure}

Total load and PV on each feeder is highlighted in figure~\ref{fig:mmgldpv}.
Since, these feeders are adjacent to each other, the overall PV profile looks
similar, but the load profiles are different to highlight the difference in
characteristics of these feeders. The overall load is considerable lower on
day-2 in feeder-2. With 4 MW of PV in each feeder, it would be challenging to
absorb all the PV in feeder-2 with the smaller 2 MW battery.

\iffalse
\begin{figure*}[t]
  \centering
  \includegraphics[width=0.7\textwidth]{_figs/19_day1_topo.png}
  \caption{Topology changes between Microgrid 1 \& 2 on \textbf{Day-1} of restoration. (\textit{top}) Total load and PV
  profile on microgrid 1 \& 2 with topology identification number during different time periods. (\textit{left bottom})
  More loads connected to MG-1 when no PV is available in the network. (\textit{right bottom}) Load group 10 connected
  to MG-1 instead of MG-2 in exchange for load group 6}
  \label{fig:mmgday1}
\end{figure*}

\begin{figure*}[t]
  \centering
  \includegraphics[width=\textwidth]{_figs/19_day2_topo.png}
  \caption{Topology changes between Microgrid 1 \& 2 on \textbf{Day-2} of restoration. (\textit{top}) Total load and PV
  profile on microgrid 1 \& 2 with topology identification number during different time periods. (\textit{bottom}) More
  load groups are switched from MG1 to MG2 due to low load on MG2, hence more probability of service.}
  \label{fig:mmgday2}
\end{figure*}
\fi

Table~\ref{tab:mmgtopochanges} highlights the topology changes on day-1 and
day-2 of the restoration. The change in topology on day-1 is minimal since the
load and PV are quite similar between the two microgrids. Initially, more load
groups are connected to MG1 since it has higher sized battery. During PV, load
group 10 which is a critical load group is connected to MG1 in exchange for load
group 5 which is a non-critical load gorup.

On day-2 we see a lot of topology changes that happen towards the end of
restoration which is due to the limited resources towards the end of
restoration. Also, more load groups are shifted from MG1 to MG2 eventhough it
has a smaller battery size because the peak load on day-2 in MG2 is much lower
than MG1, so MG1 load groups have more chance of being served when connected to
MG2.

\begin{table*}[t]
  \centering
  \caption{Optimal Topology Changes on Day-1 and Day-2}
  \label{tab:mmgtopochanges}
  \begin{tabular}{cccl}
  \hline
  \multicolumn{1}{c|}{\textbf{Duration}} & \multicolumn{1}{c|}{\textbf{MG1 Load Groups}} & \multicolumn{1}{c|}{\textbf{MG2 Load Groups}}                                        & \multicolumn{1}{c}{\textbf{Comments}}                                                                                                                                                  \\ \hline
  \hline
  \multicolumn{4}{c}{\textbf{Day-1}}                                                                                                                                                                                                                                                                                                                                     \\ \hline
  \multicolumn{1}{c|}{00:00 to 12:00}    & \multicolumn{1}{c|}{1, 2, 3, 4, 5, 6}         & \multicolumn{1}{c|}{7, 8, 9, 10}                                                     & LG-6 connected to MG1 due to bigger battery size.                                                                                                                                      \\ \hline
  \multicolumn{1}{l|}{12:00 to 15:00}    & \multicolumn{1}{c|}{1, 2, 3, 4, 10}           & \multicolumn{1}{c|}{5, 6, 7, 8, 9}                                                   & \begin{tabular}[c]{@{}l@{}}Since 2MW battery in MG2 cannot absorb all PV, \\ might lead to outage of LG-10.\end{tabular}                                                               \\ \hline
  \multicolumn{1}{c|}{15:00 to 24:00}    & \multicolumn{1}{c|}{1, 2, 3, 4, 5, 6}         & \multicolumn{1}{c|}{7, 8, 9, 10}                                                     & LG-6 connected back to MG1 due to bigger battery size.                                                                                                                                 \\ \hline
  \multicolumn{4}{c}{\textbf{Day-2}}                                                                                                                                                                                                                                                                                                                                     \\ \hline
  \multicolumn{1}{c|}{00:00 to 12:00}    & \multicolumn{1}{c|}{1, 2, 3, 4, 5, 6}         & \multicolumn{1}{c|}{7, 8, 9, 10}                                                     & \begin{tabular}[c]{@{}l@{}}More chance of load being served in MG1 since it has\\ higher battery size. Hence, LG6 connected to MG1.\end{tabular}                                       \\ \hline
  \multicolumn{1}{l|}{12:00 to 15:00}    & \multicolumn{1}{c|}{1, 2, 3, 4, 5}            & \multicolumn{1}{c|}{6, 7, 8, 9, 10}                                                  & Default topology of feeder is maintained.                                                                                                                                              \\ \hline
  \multicolumn{1}{c|}{15:00 to 18:00}    & \multicolumn{1}{c|}{1, 2, 3, 4, 5, 6}         & \multicolumn{1}{c|}{7, 8, 9, 10}                                                     & More chance of service to load groups in MG1.                                                                                                                                          \\ \hline
  \multicolumn{1}{l|}{18:00 to 21:00}    & \multicolumn{1}{c|}{1, 2, 3, 4}               & \multicolumn{1}{c|}{5, 6, 7, 8, 9, 10}                                               & \multirow{2}{*}{\begin{tabular}[c]{@{}l@{}}More loads connected to MG2 even though lower\\ battery size because loading is much lower in MG2\\ compared to MG1 on day-2.\end{tabular}} \\ \cline{1-3}
  \multicolumn{1}{l|}{21:00 to 24:00}    & \multicolumn{1}{c|}{1, 3, 4}                  & \multicolumn{1}{c|}{\begin{tabular}[c]{@{}c@{}}2, 5, 6, 7, 8, \\ 9, 10\end{tabular}} &                                                                                                                                                                                        \\ \hline
  \end{tabular}
\end{table*}

\begin{figure}[htpb]
  \centering
  \includegraphics[width=3.5in]{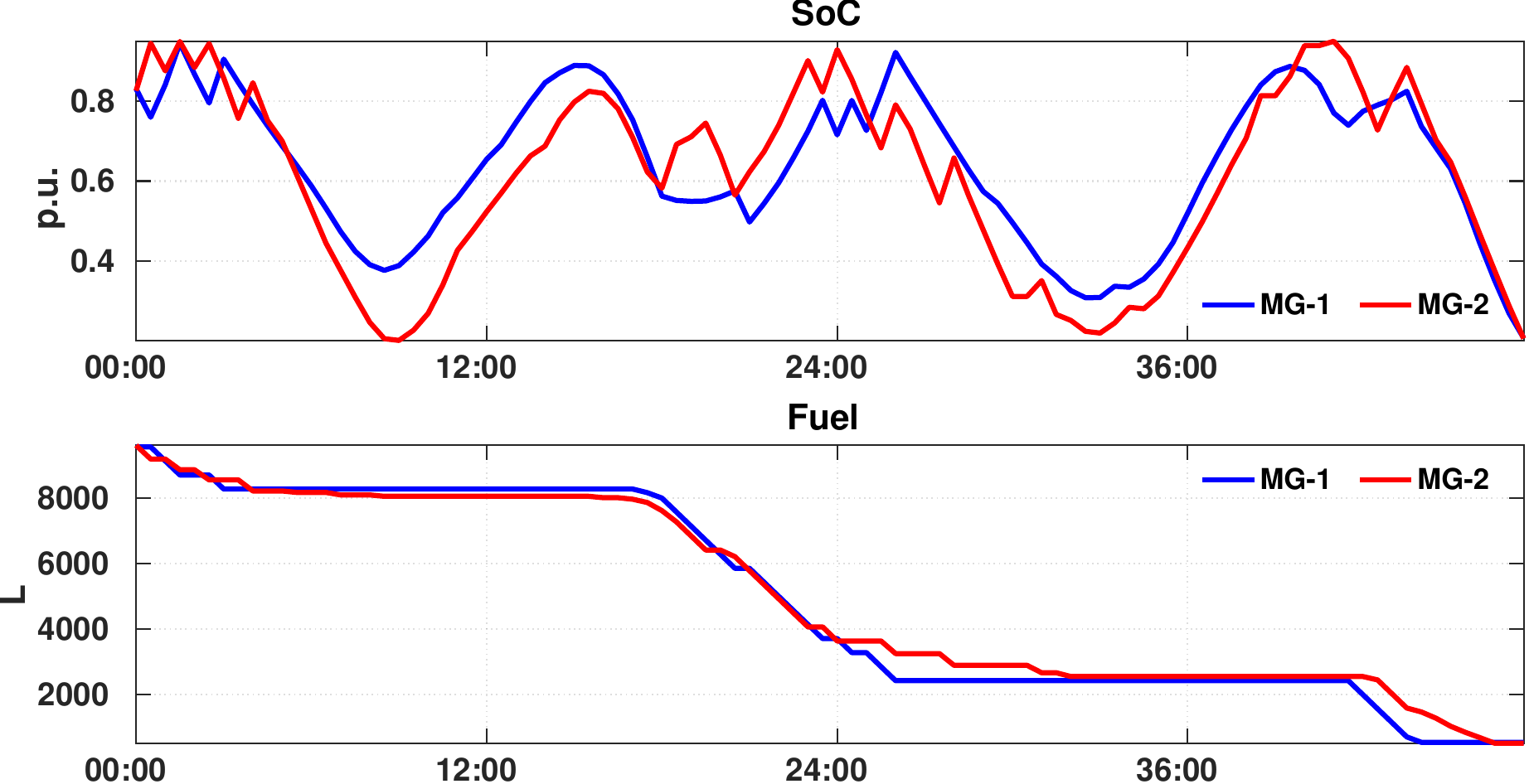}
  \caption{(\textit{top}) State of Charge of Mobile Energy Storage Devices and (\textit{bottom}) Fuel usage of Diesel
  Generators during the two day restoration period.}
  \label{fig:mmgsocfuel}
\end{figure}

The SoC of mobile energy storage devices and the fuel of diesel generators in
both microgrids are highlighted in figure~\ref{fig:mmgsocfuel}. Both battery and
DG have been appropriately used and completely depleted by the end of the
restoration. The devices were kept within limits even with significant number of
topology changes.

\begin{figure}[htpb]
  \centering
  \includegraphics[width=3.5in]{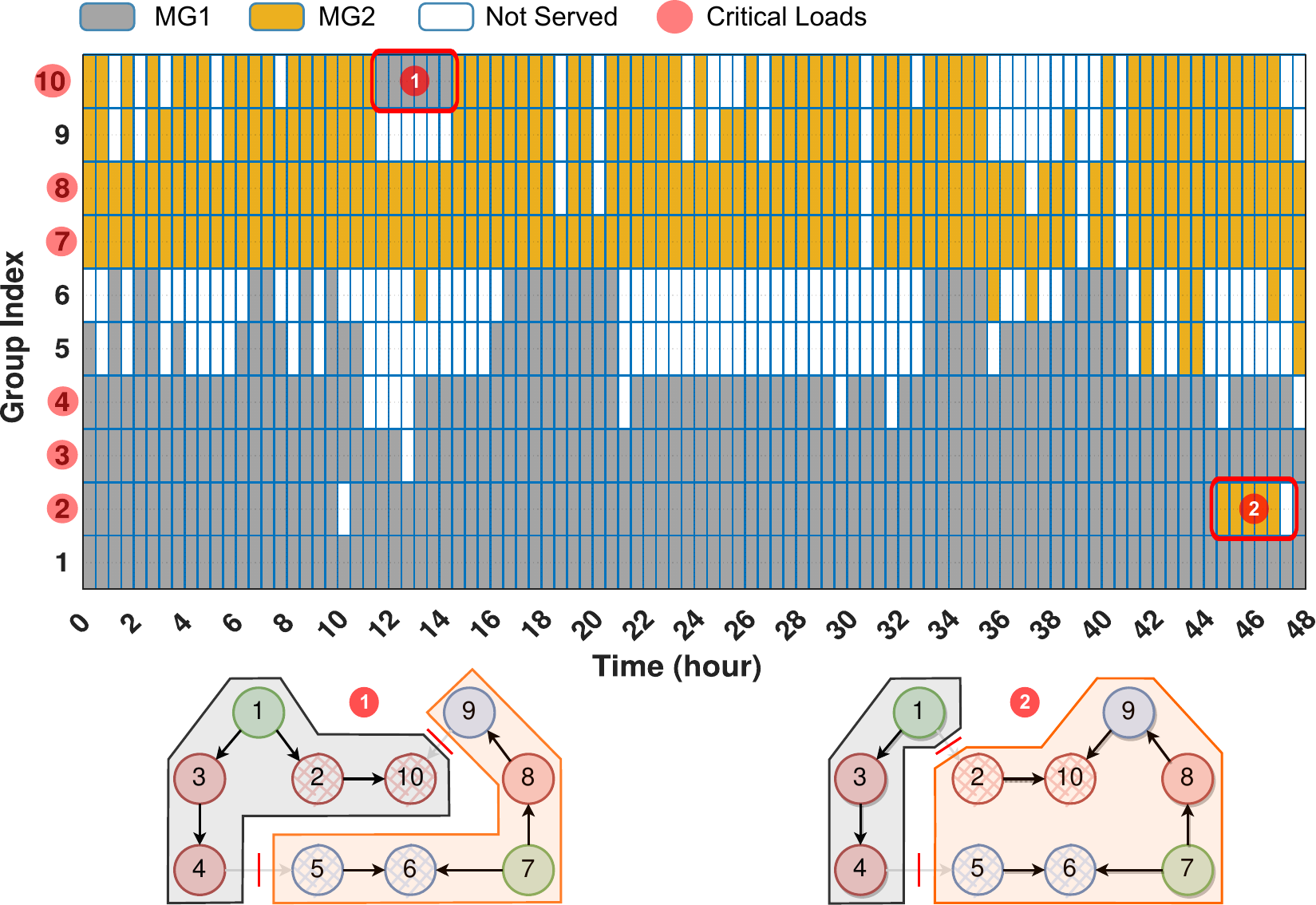}
  \caption{Load Group connectivity status highlighted using different colors to indicate which microgrid was utilized to
  energize the load groups. Two topology important topology changes to improve critical load connectivity is highlighted
  in red with corresponding topologies.}
  \label{fig:mmglgserved}
\end{figure}

The load group connectivity status is highlighted in figure~\ref{fig:mmglgserved}
which shows how different load groups are served by the two microgrids in the
system. There are instances where load groups are not served by the microgrids
even though they are assigned to the microgrid by multi-microgrid algorithm. The
downside of multi-microgrid algorithm is that when a load group is assigned to
a particular microgrid there is no guarantee that the load group will be served
by that microgrid which is subject to availability of resources and handled by
the individual energy management scheme. Nevertheless, we do minimize such
outage of critical loads which can be communicated to the customer in advance
and the customer can prepare back up local generation during these scheduled
outage hours.

Critical load groups 4 and 10 have significant duration of unserved periods by
both microgrids and this is because under any topology configuration the
distance from load group 4 and 10 to microgrid resources 1 and 7 is long. Hence,
to serve these load groups all the other load groups between the resources and
these groups need to be served which is challenging with limited resources and
during peak load duration. Hence, load groups 4 and 10 have maximum number of
unserved duration.

To analyze the performance of our proposed multi-microgrid algorithm we compare
the results against base case where the topology of the microgrids are fixed
throught the restoration period. Nodes 1 through 5 are always connected to MG1
and 6 through 10 to MG2 through out the 2 day restoration.

\begin{figure}[htpb]
  \centering
  \includegraphics[width=3.5in]{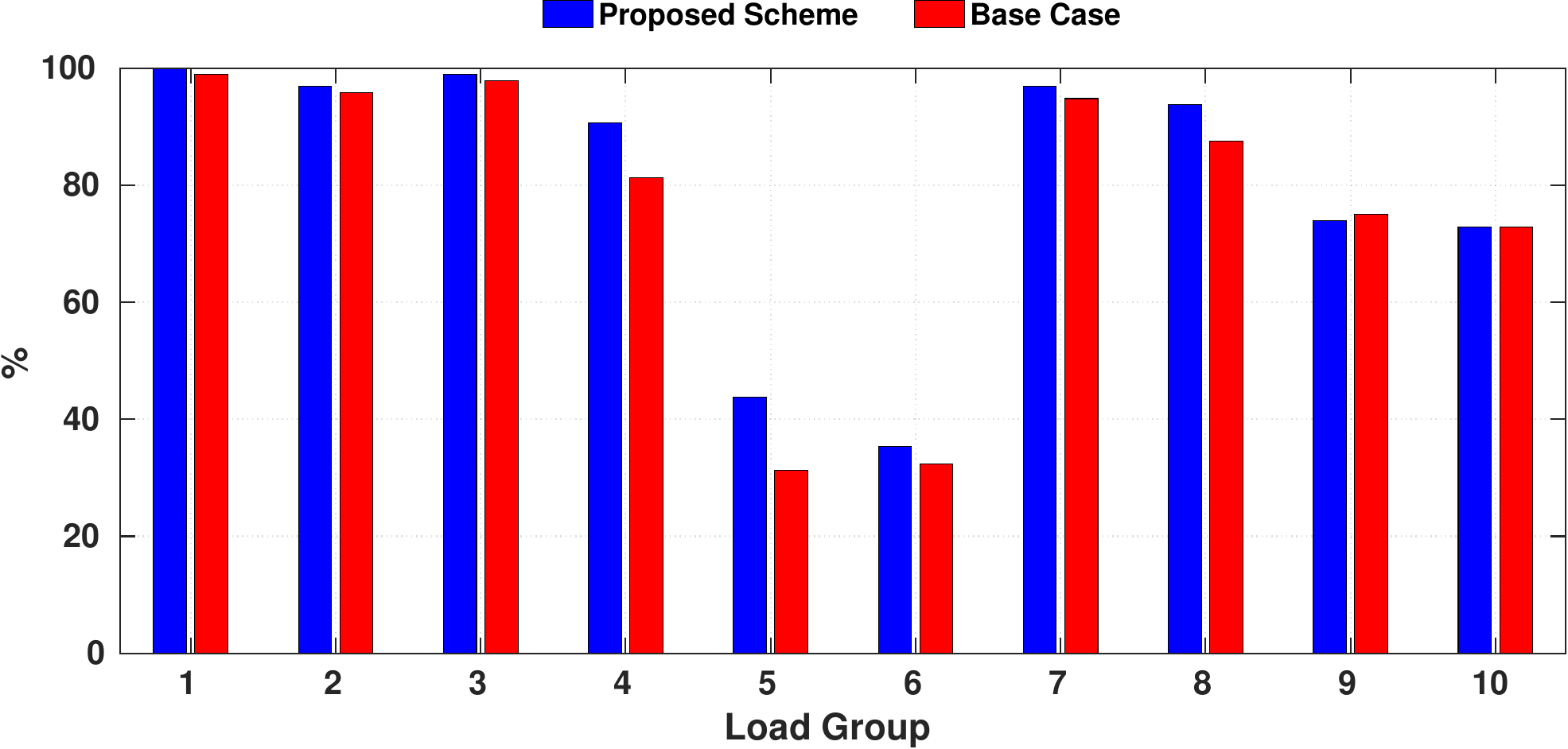}
  \caption{Percentage of Load Groups served during restoration period between base case (fixed topology) and proposed
  scheme (flexible topology). Critical Load Groups: 2, 3, 4, 7, 9, 10.}
  \label{fig:mmglgper}
\end{figure}

Percentage service to each load group is compared between proposed scheme and
base case in figure~\ref{fig:mmglgper}. It can be seen that service to critical
loads is significantly increased with proposed scheme and service to leaf nodes
like load groups 5 and 6 is also considerably increased. Due to efficient
management of the leaf nodes in flexible topology, other nodes pertaining to
fixed topology like load group 4 and 8 also have improved service. This
highlights the advantage of coordinating multi-microgrids during restoration
rather than using a fixed topology. The overall PV utilization with proposed
scheme is \textbf{82}\% which is higher than the utilization by base case with
\textbf{79.12}\%.

\section{Conclusion}
This paper highlights the advantages of optimizing resource coordination across
multiple feeders during prolonged outages through dynamic multi-microgrid
formation. We introduce a graph-theory based multi-microgrid formation problem,
extending it to multiple feeders with adjustable lateral lengths to mitigate
potential voltage issues. A case study, using a synthetic two-feeder system with
varied load characteristics derived from a realistic IEEE-123 node system,
demostrates substantial enhancements in service to critical and non-critical
loads. Coordinating microgrid resources across multiple feeders also results in
improved utilization of photovoltaic (PV) systems compared to a fixed microgrid
topology.

\bibliography{MyLibrary}
\bibliographystyle{IEEEtran}

\end{document}